\documentclass{article} 
\usepackage{graphicx}

\def\be{\begin{equation}}
\def\ee{\end{equation}}
\def\bea{\begin{eqnarray}}
\def\eea{\end{eqnarray}}

\newcommand\noi{\noindent}
\newcommand\etal{{\it et al}.\ }

\newcommand\epe{\epsilon^\prime/\epsilon}
\newcommand\ep{\epsilon}
\newcommand\sintb{\sin2\beta}
\newcommand\rhob{\bar \rho}
\newcommand\etab{\bar \eta}
\newcommand\md{\Delta m_{B_d}}
\newcommand\ms{\Delta m_{B_s}}
\newcommand\bbbar{B - \bar B}
\newcommand\bsbsbar{B_s - \bar B_s}

\newcommand\bpsik{B \to \psi K_s}
\newcommand\vtdts{V_{td}/V_{ts}}

\newcommand\bra{\langle}
\newcommand\ket{\rangle}

\def\bkhat{\hat B_K} 

\begin{document}
\rightline{BNL-HET-03/14}
\begin{center}
{\large\bf Lattice Matrix Elements and CP Violation in B and K Physics: Status and 
Outlook$^*$}
\vspace{0.2in}

Amarjit Soni\\ 
\noindent High Energy Theory Group, Physics Department\\
Brookhaven National Laboratory,Upton, NY 11973\\ 
(email: soni@bnl.gov)
\end{center}
\footnotetext{Invited talk at 9th International Symposium on Particles, Strings and Cosmology (PASCOS 03), Mumbai
(Bombay) India, 3-8 Jan 2003}
\begin{quote} 
\begin{center}
ABSTRACT
\end{center}

Status of lattice calculations of hadron matrix elements
along with CP violation in B and in K systems is reviewed.
Lattice has provided useful input which, in conjunction with
experimenatl data, leads to the conclusion that CP-odd
phase in the CKM matrix plays the dominant role in the
observed asymmetry in $B \to \psi K_s$. It is now
quite likely that any beyond the SM, CP-odd, phase
will cause only small deviations in B-physics. Search 
for the effects of the new phase(s) will consequently require
very large data samples as well as very precise theoretical
predictions. Clean determination of {\it all} the angles
of the unitarity triangle therefore becomes essential.
In this regard $B \to K D^0$ processes play a unique
role. Regarding K-decays, remarkable progress made by theory with regard
to maintenance of chiral symmetry on the lattice is
briefly discussed. First application already provide 
quantitaive information on $B_K$ and the $\Delta I=1/2$
rule. The enhancement in $Re A_0$ appears to arise solely
from tree operators, esp. $Q_2$; penguin contribution to $Re A_0$
appears to be very small. However, improved calculations
are necessary for $\epe$ as there the contributions
of QCD penguins and electroweak penguins largely seem to cancel.
There are good reasons, though, to believe that these
cancellations will not survive improvements
that are now underway. Importance of determining the unitarity
triangle purely from K-decays is also emphasized. 
\end{quote}
\section{Introduction}

With important input from the lattice along with the classic results of indirect CP 
violation in $K_L \to \pi \pi$, the asymmetric B-factories with measurements of CP
asymmetry in $B \to \psi K_s$ are providing valuable support to the CKM paradigm 
of CP violation\cite{ckm}. It is now clear that the CP-odd phase in the CKM matrix
is the dominant source of CP violation in $B \to \psi K_s$. However, as is well known
essentially compelling theoretical arguements suggest that new
CP-odd phase(s) should exist due to physics beyond the SM (BSM). At the same time
there is no good reason to think that their effects in B-physics would be 
particularly large.
Indeed, SM teaches a valuable lesson in this regard:
even though the CKM phase causes a huge asymmetry (i.e. $O(1)$) in $B \to \psi K_s$,
its effects in CP violation in K-decays is miniscule $\approx 10^{-3}$. 
Clearly this realization should motivate us to prepare for
small deviations from the predictions of the SM in B-Physics even if the new CP-odd
phase is large.   
For this reason, not only we need very large data samples of B's giving impetus to Super-B factories
along with BTeV and LHCB, we also need exteremely precise tests of the SM.
Residual theory errors are a serious cause of concern as they can easily thwart
experimental efforts for search of BSM CP-odd phase(s)\cite{pristine}.   

With this perspective in mind, a  brief discussion of the lattice method and results for the
hadronc matrix elements that are important for weak interaction
phenomenology are given in Sections I to IV. 
Therein I also discuss  some of the anticipated experimental input that could
help attain greater precision in constraining CKM parameters.  
Section V emphasizes concern about residual errors in theory and 
the importance of clean determinations, i.e. without theoretical 
assumptions, of all the angles
of the unitarity triangle. In this regard the special role of $B \to K D^0$ processes is also
emphasized there.

Progress made in the past few years with regard to maintenance of exact chiral
symmetry on the lattice is outlined in Section VI along with application
of this development to $B_K$.    

Section VII gives a brief report on the results from the 1st application 
of domain wall fermion method, which exhibit excellent chiral behavior,
to $K \to \pi \pi$, $\Delta I=1/2$ rule and $\epe$.
These 1st applications give good insight to the $\Delta I=1/2$ rule; in particular,
contribution of penguin operators to $ReA_0$ in our lattice calculations appears to be extremely
small and most of the enhancement seen in $ReA_0$ is originating
from the tree operator, $Q_2$. Unlike in the case of the $\Delta I=1/2$ rule,
the approximation currently used though appear too crude to give reliable
information on $\epe$. This difficulty arises as contributions of QCD penguins
and electroweak penguins substantially cancel. 
However, there are very good reasons to suspect that this cancellation 
is not ``natural" and is unlikely to survive 
as calculations are improved.

For the purpose of stringent tests of the CKM model of CP violation
a separate determination of the unitarity triangle 
purely from K-decays, to be compared to that obtained from B-physics,
is highly desirable and this is finally emphasized in Section VIII.

\section{Lattice Methodology: a very brief recapitulation}

Recall, Green's functions are calculated by numerical evaluation of the
Feynman path integral.

\begin{equation}
\langle 0|Q|0\rangle =\frac{\int~DU~W(Q)_M det~M(U)exp[-S_g(U)]}{\int~DU~ 
det~M(U)exp[-S_g(U)]}
\end{equation}

As it stands, dependence of the quark matrix M in this expression
on the link variables renders its evaluation extremely difficult.
To fecilitate the numerical calculation, one often uses the quench
approximation (QA)and sets $detM = 1$. Physically, this approximation 
corresponds to neglect of the $q \bar q$ vacuum polarization loops in the 
propagation of the gauge field. The hint that this may be a reasonable 
approximation originally came from deep inelastic scattering experiments
wherein the effect of $q \bar q$ pairs in the ``sea" is accurate to
about 15\% \cite{dwein}. There are, though, very good reasons that tell us that 
the
accuracy of the QA in lattice computations is process dependent.

In the past several years more and more ``unquenched" simulations,s
i.e. those not using the QA so that dynamical $q \bar q$ pairs are
included, have been underway.  These studies show that QA seems to be valid to 
about $5-10\%$ accuracy in non-singlet hadron spectrum \cite{dtous}. 


On the other hand, dynamical quarks seem to increase the B-meson pseudoscalar
decay constant quite appreciably, (at least when $m_\rho$ is used to set the 
scale) \cite{heplat} 
\begin{equation}
f_B^{N_f=3}/f_B^{N_f=0} = 1.23 \pm 0.04 \pm 0.06 
\label{eq:fbnf3ratio}
\end{equation}

In addition to the QA there are several other sources of systematic
errors in a typical lattice gauge calculation. Chief among these are finite
(box) size and finite lattice spacing (a) errors. Also most lattice simulations
are done with rather large values of masses of light (u,d,s) quarks and rather low 
values for the  b quark mass, compared to their physical values. Painstaking and 
elaborate efforts become necessary to accurately extract from the data 
information relevant to the physical case. This may, for example, require 
extrapolation of the data (at a fixed gauge coupling, or lattice spacing)
as a function of quark mass to the chiral limit and also extrapolation
of the data as a function of the lattice spacing to the continuum limit (i.e. 
lattice spacing goes to zero).
Furthermore, simulations for a fixed gauge coupling at two or more volumes are 
often needed for extrapolation to the infinite volume limit.

\subsection{Some Examples of Brute-Force}

Relevant to this talk there are three works which serve to illustrate
computational brute force used in bringing them to fruition; these are
$B_K$, $f_B$ and $K -> 2 \pi$.

\begin{enumerate}

\item $B_K$: A major accomplishment of the lattice gauge effort, and in
particular of the JLQCD group is their result\cite{aokione,kilcup}, $\bkhat = 
0.860 \pm 
0.058$ in the QA.  During the first 6-7 years ('84-'91 ), several exploratory 
attempts were made\cite{cbas88}.
Methodology was in place around '91 \cite{ssharp} which was followed by several 
years of
intensive computations leading to the final result obtained around '98. In the 
past few years this important result has been the 
focus of further checks and 
confirmation using other fermion discretizations, i.e. 
Wilson \cite{becir} as well as the 
newer discretizations: domain wall fermions \cite{tblum,alikh,blumone} and the 
overlap fermions \cite{milc,ngarr}.  The 
results of these methods are in rough agreement with the JLQCD result; however, 
with domain wall quarks (DWQ) method the central value of $B_K$ tends to be 10-15\% below 
the JLQCD result which may amount to a discrepancy of around 1-2 $\sigma$. 
(see Fig.~\ref{bkfig})
More 
precise calculations with these newer discretizations including a study with 
dynamical domain walls \cite{izubuchi} is now underway.

\begin{figure}
\includegraphics[scale=0.6]{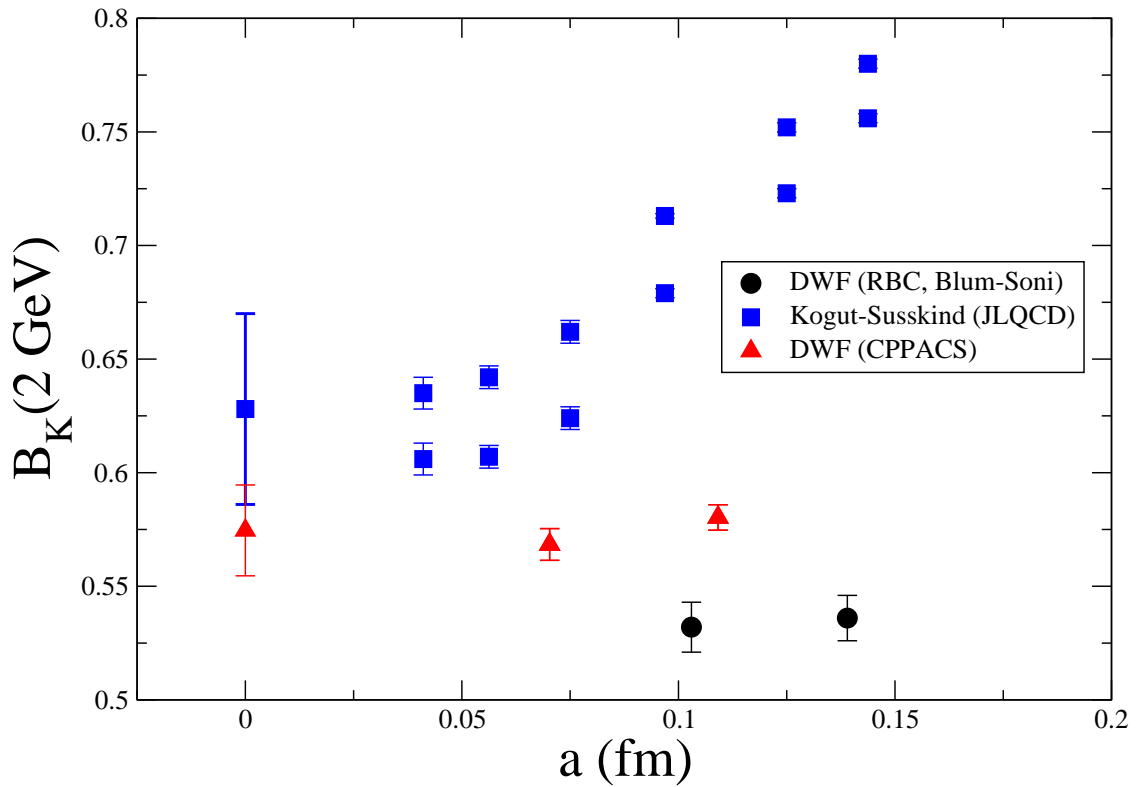}
\caption{
$B_K$ vs. a, the lattice spacing. Only the 2 data points (circles)
from DWF use non-perturbative renormalization of lattice
operators, all others use one-loop perturbation theory. Of
the circled DWF 2 points, the coarser spacing (right)
corresponds to data from NERSC whereas 
the finer one (left) belongs to RBC collaboration \cite{blumone}} 
\label{bkfig}
\end{figure}

\bigskip

\item $f_B$:  Another example is provided by $f_B$, the B-meson
pseudoscalar decay constant wherein the ``heavy" b-quark mass $\approx$ 4.5 GeV 
represents an additional technical problem.
After the initial 5-6 years of exploratory works the computational
strategy became quite well known around '92 \cite{bls}. Indeed the result in 
the QA has 
been quite stable and withstood checks with the use of different techniques 
\cite{nyam,nucl}.
In the past few years there has been some weak indication from
experiment that heavy-light decay constant are somewhat smaller in the QA 
compared to experiment \cite{hepph}(i.e. full QCD).
Indeed after years of persistent study the MILC collaboration has now
finished the calculation of $f_B$ in quenched as well as in full QCD
(i.e. with three light flavors of dynamical quarks) \cite{heplat,co94} and find, 
the ratio given in eq.(2).  

giving $f_B \simeq 207 \pm 35$ MeV\cite{error}.
Clearly many independent checks and confirmations will take place in the next 
few years.

\item $K \to \pi \pi$ and $\epsilon'/\epsilon$. Our third example is
the calculation of the matrix elements of $K \to 2\pi$ and $\epsilon'/\epsilon$ 
using domain wall quarks (DWQ) by the CP-PACS \cite{noaki}and RBC collaborations 
\cite{blumone}.
Unlike $B_K$ and $f_B$ this is a first attempt to address $K\rightarrow 
2\pi$ by both the 
collaborations in which not only QA but also a few other key approximations are 
made [see below].
Nevertheless, given the complexity of the problem it must be considered an  
inportant accomplishment which even at this early stage is providing very useful information on 
the long standing issue of the $\Delta I=1/2$ rule. However, its repercussions 
on $\epsilon'/\epsilon$ require careful study of systematic errors and improved 
calculations, which could take another few years.

\end{enumerate}

\section{Lattice Matrix Elements and CKM Constraints}

In the Wolfenstein representation, the CKM matrix can be parameterized
in terms of the four parameters, $\lambda$, A, $\rhob$ and
$\etab$ \cite{hep0101336}. Of these $\lambda = \sin \theta_c = 0.221\pm 
0.002$,
is the best known, A is known with modest accuracy, $A=0.847\pm~041$ and 
$\bar\rho$ and $\bar \eta$ are poorly known.
An important objective where lattice can help is in determination of
$\bar \eta$ and $\bar \rho$ accurately. $\bar \eta$ is intimately
related to the CKM phase $\delta_{13}$ \cite{kpart}; indeed SM cannot 
accommodate any CP
violation if $\bar \eta = 0$.

The basic strategy is very simple. Assuming the SM is correct, and using
the necessary theoretical input one translates experimental results on to an
allowed domain on the $\bar \eta$ - $\bar \rho$ plane. If a 
(new) experimental result requires value of $\bar\rho $ and/or $\bar\eta$ that are 
inconsistent with those extracted from existing experiments then that could mean 
a failure of the SM.

\subsection{Theoretical background and brief comments}

For the past several years, the following four experimental measurements
have been used for an extraction of $\bar\eta$ and $\bar\rho$:

\begin{enumerate}

\item The indirect CP violation parameter, $\epsilon = (2.274 \pm .017)\times
10^{-3}$

\item The $B_d-\bar B_d$ mass difference, $\md = 0.487 \pm 0.014 ps^{-1}$

\item The $B_s-\bar B_s$ mass difference, for which at the moment only
a lower bound exists, $\ms \ge 15.0 ps^{-1}$ at 95\% CL.       

This important bound is provided by experiments at LEP and SLD \cite{lepb}.
It is widely anticipated that an actual measurement of $\ms$ 
(rather than just a bound) will be accomplished at the Tevatron in the next few 
years.  This will be very important for CKM determinations as the 
ratio $\md/\ms$ can give $V_{td}/V_{ts}$ if the SU(3) 
breaking ratio of hadronic
mixing matrix element could be determined\cite{bern}.

\item $R_{uc} \equiv b \to u l \nu / b \to c l \nu = 0.085 \pm 0.017.$
\end{enumerate}

Recall\cite{buch} 
\begin{equation}
|\epsilon| = \tilde{M}_{K} C_K A^2\lambda^6 {\etab}\{\eta_1 S(x_c) + 
\eta_2 S(x_t)A^2\lambda^4 (1 - {\rhob}) + \eta_{3} S(x_c,x_t)\}
\end{equation}



Here $x_q = m^2_q/M^2_W$, where $q = u,c,t$ i.e. the virtual quarks in the 
box graph for $K^0-\bar K^0$ oscillations, and $S(x_q)$ are the so called Inami-Lin 
functions \cite{tinam}. Also,

\begin{equation}
C_K = \frac{G^2_F M^2_W m_K}{6\sqrt 2 \pi^2 \Delta m_K}
\end{equation}

\begin{equation}
\tilde{M}_K =\frac{3}{8} M^{2}_K \langle K^{o}|[\bar{s}\gamma_\mu(1-
\gamma_5)d]^2|\bar {K}^{o}\rangle
\end{equation}

\noi is essentially the hadronic matrix element. Once the $\tilde{M}_K$ is known 
$\etab, \rhob$ can be constrained through the 
use of eq.3. This matrix element is often parametrized in terms of $B_K$ which 
should equal 1 if vacuum saturation approximation (VSA) holds. Since $f_K$ (and 
$m_K$) is known quite precisely from experiment, evaluation of the  matrix 
element is completely equivalent here to that of $B_K$.

Similarly, we note that for $B_d -\bar B_d$ oscillation

\begin{equation}
x_{Bd} =\tilde{M}_{Bd}~C_{Bd}[{(1-\bar{\rho})~^2 + \bar{\eta}}~^2]\eta_{QCD} 
S(x_t)~A^{2}\lambda^{6}/\tau_{Bd}
\label{xbd}\end{equation}

\noi where $x_{Bd}\equiv\frac{\Delta~m_{Bd}}{\Gamma_{Bd}}$ and 
$C_{Bd}=\frac{G^{2}_{F} 
m^{2}_W}{6\pi^{2}~m^{3}_{Bd}}$
Again once the hadronic matrix element, $\tilde{M}_{bd}$, is known eq. 
($\ref{xbd}$)   
can be used to 
constrain $\rhob$, $\etab$.  

This matrix element is a 3-point function, which is directly calculable on the 
lattice. More often than not, though, in analogy with the kaon case, $M_{bd}$ is 
parametrized in terms of a ``B-parameter" defined as

\begin{equation}
B_{Bd}=\frac{<B_{d}\vert[\bar{b}\gamma_{\mu}(1-\gamma_{5})d]^{2} 
\vert\bar{B}_d>}{\frac{8}{3}\tilde{m}_{Bd}^2f^2_{Bd}}
\end{equation}
Then the physical quantity $x_{Bd}$ requires both $B_{Bd}$ and $f_{Bd}$
since
the latter is not yet known from experiment. Besides since
$\tilde{M}_{Bd}$
seems to
scale roughly as $f^{2}_{Bd}$ one needs to know $f_{Bd}$ rather
accurately.
Also,
in practice in most calculations of $f_{Bd}$ one tries to fit the light
quark mass dependence through some linear function; such a fit, though
is
unlikely
to give precisely the dependence on light quark mass for $f^2_{Bd}$.

Once $B_s - \bar B_s$ oscillation are experimentally detected and $\ms$
becomes
known, then the ratio

\begin{equation}
\frac{\md}{\ms} = \frac{|V_{td}|^2}{|V_{ts}|^2}\frac
{\langle B_{d}|[\bar{b}\gamma_\mu(1-\gamma_5)d]^2|\bar{B}_d\rangle}
{\langle
{B_{s}}|[\bar{b}\gamma_\mu(1-\gamma_5)s]^2|\bar{B}_s\rangle}
\end{equation}


\noi can be used to determine $\vert~V_{td}\vert$ if the ratio of hadronic matrix 
elements could
be determined from the lattice.

Since this ratio of matrix elements is completely dependent on SU(3)
breaking effects $(s\leftrightarrow d)$ it is expected to be close to unity.
The objective of lattice calculations should be a precise evaluations
of this SU(3) breaking and this necessitates an accurate treatment of light 
quarks.

Again, introducing ``B-parameters" for $B_d$ and $B_s$ mesons we can rewrite

\begin{equation}
\frac{x_{Bd}}{x_{Bs}} = 
\frac{\tau_{Bs}}{\tau_{Bd}}\frac{m_{Bd}^2}{m_{Bs}^2}\frac{1}{\xi^2}~
\lambda^2[(1-
\rhob)^2 + 
\etab^2]
\end{equation}

where $\xi$
is the SU(3) breaking ratio, 

\begin{equation}
\xi = \frac{f_{Bs}}{f_{Bd}}\sqrt\frac{B_{Bs}}{B_{Bd}}
\end{equation}

Finally, the semi-leptonic branching ratio $b \to u l \nu /b \to c l
\nu$

is another important way of constraining $\rhob , \etab$ as it is a
function of $V_{ub}/V_{cb}$,

\begin{equation}
\frac{\vert~V_{ub}\vert}{\vert~V_{cb}\vert}= \lambda(\rhob^{2}+\etab^{2})/(1-
\frac{\lambda^{2}}{2})
\end{equation}

To deduce  $V_{ub}/V_{cb}$, from the experimental measurement of the 
branching ratios requires corresponding form factors for exclusive reactions 
wherein lattice methods can be useful \cite{aelk}.  In the interest of brevity, 
we will not 
disuss this here.

\section{Lattice Input for CKM Fits}

Table~\ref{ckm-fits}  shows the input from the lattice, experiment and elsewhere
used by us \cite{atwoodone} and compare it with the works of Ciuchini et al 
\cite{ciuch} and Hocker et al \cite{hocker}.  
The corresponding determination of the 
CKM parameter $\rhob, \etab$ and unitarity angles $\alpha$, $\beta$, $\gamma$ as 
well as several other quantities of interest are also shown. Note that
our error on $f_{Bd} \sqrt B_{Bd}$ and on $\xi$ are appreciably bigger than used 
in the other studies. This is especially so for $\xi$, where for quite sometime we 
have been cautioning that the error of $\approx 0.05$ that was commonly taken was a 
serious underestimate \cite{intmor}.  Recently Kronfeld and Ryan 
\cite{aknon} and Yamada \cite{nyam} have also 
argued for a reassessment of errors on $\xi$ due to the presence of
chiral logs.  Following this development as of LAT'02 larger error on $\xi$ is 
now being widely advocated; for example, Lellouch \cite{llellh} in his review at 
ICHEP02 
summarized $\xi = 1.18 \pm 0.04 ^{+12}_{-0}$.

\begin{table}[t]
\caption{Comparison of some fits.\label{tabone}}
\hspace*{-.5in}\begin{tabular}{|c|c|c|c|}
\hline
Input Quantity
& Atwood \& Soni\cite{atwoodone}
& Ciuchini {\it et al}\cite{ciuch}
& Hocker {\it et al}\cite{hocker}
\\
\hline
\hline
$R_{uc} \equiv |V_{ub}/V_{cb}|$ & $.085\pm.017$ & $.089\pm.009$ &
$.087\pm.006\pm.014$ \\
$F_{B_d} \sqrt{\hat B_{B_d}}$ MeV & $230\pm50$ & $230\pm25\pm20$ &
$230\pm28\pm28$ \\
$\xi$ & $1.16\pm.08$ & $1.14\pm.04\pm.05$ & $1.16\pm.03\pm.05$ \\
$\hat B_K$ & $.86\pm0.15$ & $.87\pm0.06\pm0.13$ & $.87\pm.06\pm.13$
\\
\hline
Output Quantity & & & \\
\hline
$\sin2\beta$ & $.70\pm.10$ & $.695\pm.065$ & $.68\pm.18$ \\
$\sin2\alpha$ & $-.50\pm.32$ & $-.425\pm.220$ & \\
$\gamma$ & $46.2^\circ\pm9.1^\circ$ & $54.85\pm6.0$ & $56\pm19$ \\
$\bar \eta$ & $.30\pm.05$ & $.316\pm.040$ & $.34\pm.12$ \\
$\bar \rho$ & $.25\pm.07$ & $.22\pm.038$ & $.22\pm.14$ \\
$|V_{td}/V_{ts}|$ & $.185\pm.015$ & & $.19\pm.04$ \\
$\Delta m_{B_s}(ps^{-1})$ & $19.8\pm3.5$ & $17.3^{+1.5}_{-0.7}$ &
$24.6\pm9.1$ \\
$J_{CP}$ & $(2.55\pm.35)\times10^{-5}$ & & $(2.8\pm.8)\times10^{-5}$ \\
$BR(K^+\to\pi^+\nu\bar\nu)$ & $(0.67\pm0.10)\times10^{-10}$ & &
$(.74\pm.23)\times10^{-10}$ \\
$BR(K_L\to\pi^0\nu\bar\nu)$ & $(0.225\pm0.065)\times10^{-10}$ & &
$(.27\pm.14)\times 10^{-10}$ \\
\hline
\end{tabular}
\label{ckm-fits}
\end{table}

\bigskip  

The SM fits now give $(sin 2\beta)^{SM} = 0.70 \pm 0.10$ as well as allowed
ranges for $\gamma, \etab, \rhob$ etc (see Table~\ref{ckm-fits}). 
While these fits provide 
fairly restrictive range for $\beta$ and $\gamma$, $\alpha$ is constrained 
rather poorly. Note also that $\bsbsbar$ mass difference $\ms = 19.8 \pm 3.5 
ps^{-1}$ is 
now constrained with a one sigma accuracy of about $15\%$; measurements at the 
Tevatron and later at the LHC should be able to test this important   
prediction of the SM. Meanwhile measurements of the CP asymmetry in $\bpsik$ 
is already providing quite an impressive determination, $\sintb = 0.734
\pm 0.054$ \cite{belle} in good agreement with the theoretical prediction. It is 
important to
note also that just in the past year B-factory experiments have improved the
determination of $\sintb$ from an error of $\pm$ 0.10 down to $\pm$ 0.05.
With the anticipated increase in luminosities of the B-factories, along with 
results from the Tevatron, this error should go down further to $\approx 0.02$ 
in another year or two.
(Recall that the intrinsic theory error in the determination of $\sintb$ 
is expected to be about $\le 0.01$)\cite{ynirh}

It is instructive to reflect on the 
pace of theoretical progress in constraining 
$\sintb^{SM}$. For this purpose we may
compare the inputs used in fits of $\approx$ 1995 
\cite{ashlat}, with 
that of 
$\approx$ 2001\cite{atwoodone}.
Indeed the 2001 fit has reduced the 
error on $\sintb$ from 0.20 to 0.10; correspondingly the error on $\etab$ 
and on 
$\ms$ is also appreciably reduced. However, only some of this improvement can
be related directly to lattice computations. In fact it seems
that a large portion of the improvement is due actually to the reduction 
especially in 
the error on $V_{cb}$ and to some degree on $V_{ub}/V_{cb}$ wherein the role
of the lattice is less clear.

What should be clear is that it will be extremely difficult to reduce
the theory error on $(\sintb)^{SM}$ from the current level of $\pm$ 0.10 down to 
level of 0.02
that the experiment is anticipated to reach in the very near future.  Thus to 
test the SM 
more precisely will require clean determination of the other angles $\alpha$ and 
$\gamma$ directly from experiment. We will come back to this point later on.

\subsection{Important Input from Experiment on the horizon for CKM 
Determination}
 
\begin{enumerate}

\item $B^{\pm} \to \tau^{\pm} + \nu_{\tau} (\bar \nu_{\tau})$ 

With $10^8$ or more $\bbbar$ pairs 
that BELLE and BABAR each will soon have access to, an experimental 
determination 
of $f_B$ (actually $f_B \times V_{ub})$ may be feasible. Using from the 
lattice $f_B \approx 207 \pm 35 MeV \cite{error}$  and  $V_{ub}/V_{cb} \approx 
0.085 \pm 0.017$ 
one gets an estimate, Br ($B \to \tau + \nu_{\tau} \approx (7.8 \pm 2.0) \times 
10^{-5}$).  Decays of $\tau$ into final states with $(\nu's)+ $ $\mu$ (e, $\rho$ 
or $\pi$) have a total branching ratio of around $50\%$. So with a few percent 
detection efficiency there should be a few hundreds of events for $B^{\pm} \to 
\tau^{\pm} + 
\nu_{\tau} (\bar \nu_{\tau})$, a   
respectable sample to provide a reasonable determination of $f_B\times V_{ub}$
and an important check on the lattice calculation.

\item $B \to l \nu \gamma$ 

Unlike $B \to l \nu$, $l \nu \gamma$ ($l = e, \mu$) does not suffer from helicity
suppression although it is suppressed by $\alpha$.  Emission of the photon from 
the light quark also tends to enhance the
process although precise calculation of the Br is difficult to make \cite{dga}; 
estimates \cite{ksw}
are in the range of $1 - 6 \times 10^{-6}$, i.e. about an order of magnitude
more than the 2-body helicity suppressed modes, $  B \to l \nu$.  
The constituent quark model, although too simple
to provide reliable details, perhaps does give a valid qualitative
picture indicating a ``hard" photon spectrum \cite{dga}:

\begin{equation}
\frac{dN}{d\lambda_{\gamma}}=\frac{m_{B}}{\Gamma_{l\nu\gamma}}\frac{d
\Gamma_{l\nu\gamma}}{dE_{\gamma}}=24\lambda_\gamma(1-2\lambda_{\gamma})
\end{equation}

where $\lambda_{\gamma} = E_{\gamma}/m_B$, and yields a total Br $\approx
5\times 1 0^{-6}$ with a constituent light quark mass of about 350 MeV and $f_B 
= 200 MeV$. Predictions from several other estimates are given in 
Table~\ref{lnug}. These 
radiative modes should be accessible accompanied by $\mu$ or e with the data 
samples currently available. In making contact with the phenomenological
models, energy spectra of the photon and of the neutrino [ i. e. the
invarriant mass of ($\gamma +$ the charged lepton)] would be especially
useful. Detailed experimental studies of these radiative decays would also give another handle on the 
approximate value of $f_B$.

\bigskip

\begin{table}[t] 
\caption{Sample of rough estimates 
for $B^{\pm} \to l \nu \gamma$.\label{lnug}} 
\hspace*{-.5in}\begin{tabular}{|c|c|c|}
\hline
Method & Reference & Br \\ 
\hline
const. quark model & Atwood et al\cite{dga} & $\approx 5 \times 10^{-6}$ \\ 
light cone + HQET & Korchemsky et al\cite{ksw} & $\approx 5 \times 10^{-6}$ \\ 
light-front & Geng et al\cite{ksw} & $\approx 4\times 10^{-6}$ \\
Rel. potential & Colangelo et al\cite{ksw} & $\approx 1\times 10{-6}$ \\
QCD-fact,SCET & Descotes-Genon and Sachrajda\cite{ksw} & $\approx a few\times 10^{-6}$ \\ 
\hline
\end{tabular}
\end{table}

\item $B^0 \to \rho + \gamma$

Another important input from experiment that could aid in the
determination of the CKM-parameters (esp. $\vert~V_{td}\vert/\vert~V_{ts}\vert$)  
is the reaction $B^0 \to \rho + \gamma$ 
since 

\begin{equation}
\frac{Br(B^{\circ}\rightarrow \rho^{\circ}\gamma)}{Br(B\rightarrow
K^{*}\gamma)}=\frac{[1-m^{2}_\rho/m^{2}_{B}]}{[1-m^{2}_{K^*}/m^{2}_{B}]}
\biggl[\frac{T_1^{B\rightarrow\rho}(0)}{T_1^{B\rightarrow 
K^{*}}(0)}\biggr]^2\frac{\mid 
V_{td}\mid^{2}}{\mid V_{ts}\mid ^{2}}
\label{brb}\end{equation}

The expected $Br (B^0 \to \rho + \gamma) \approx 1\times 10^{-6}$ seems
within reach of experiment.  An accurate calculation of the (SU(3)) breaking 
ratio of form factors $[T_1^{B \to \rho}(0)]/[T_1^{B \to K^*}(0)]$, for example 
by lattice methods, the feasibility of which was demonstrated long time already 
\cite{hsieh,bufor}, 
along with the anticipated experimental measurement of $B \to \rho + \gamma$ 
could lead to another determination of $V_{td}/V_{ts}$\cite{ahmed}.

This method of extracting $\vtdts$ has some advantages and some disadvantages 
compared to $\bsbsbar$ oscillations. For $B \to \rho \gamma$ the relevant 
operator is a bilinear one, whereas for $\bsbsbar$ oscillations a 4-quark
operator enters; the renormalization of a lattice 4-quark operator can be more
complicated compared to a bilinear one.  On the other hand, $B \to \rho + 
\gamma$  involves a large recoil thereby extracting the form factor at or near 
$q^2 = 0$ (where q is the 4-momentum of the photon) is numerically difficult on 
the lattice.

From the experimental side in the numerator of eq. ($\ref{brb}$) only neutral 
B's,
{$B^0,
\bar B^ 0 \to \rho^0 \gamma$} should be used as charged {$B^\pm \to \rho^\pm 
\gamma$} provides a non-negligible long-distance contribution \cite{blak} which 
is 
proportional to $V_{ub}$ (and independent of $V_{td}$) coming from the 
annihilation graph and cannot be estimated accurately.

\end{enumerate}

Due to these anticipated input from experiments along with developments in
theory, especially with the expected improvement in computational resources because  
of the Scientific Discovery Through Advanced Computing (SCIDAC) initiative \cite{scidac}, it 
is 
fairly safe to say that errors on SM parameters such as $\sintb$ will go down by 
a factor of about 2-3.  However, with larger pool of B-samples that are expected 
from B-factories and the hadron facilities, experiments should be able
to directly determine  $\sintb$ from CP asymmetry measurements in $\bpsik$
to an accuracy of 0.02; so experiment is likely to stay ahead of theory. 
Uncovering new sources of CP violation in B-physics though may well require more 
precise tests, as we will emphasize in the next few paras. 

\section{Theory Errors and the hunt for new sources of CP violation
in B-physics} 

Theory errors should be a concern as they can thwart experimental
efforts to search for the beyond the standard model (BSM)-CP odd new phase(s) 
which we will collectively denote 
as $\chi$. The main point is that $\chi$ may well cause only small deviations 
from the SM in B-physics.
Indeed the emerging  understanding of the CKM-paradigm serves as an important 
lesson in this regard.
The-CP-odd phase $\delta_{13}$, in the standard notation, ($\delta_{13}
\approx
\gamma \approx 50
\pm 10 \deg$) is not small and although it causes O(1) CP-asymmetry in
$\bpsik$ its  effects in $K_L \to \pi \pi$, $\epsilon$ or $\epsilon'$ are  
very small, $O(10^{-3} - 10^{-6})$ respectively \cite{kpart}. Analogously it is 
clearly not
inconceivable that even though a BSM-CP-odd phase $\chi$ is not small its effect 
on B-physics will be small. As an example, this may happen if $\chi$ 
arises in models with extra Higgs bosons; then its effects may be much larger in 
top physics and quite small in B-physics \cite{atwhep}.

For one thing this means we may need very large data samples of B's.
Indeed for an asymmetry of $O(10^{-3})$ (as in $K_L$), since the relevant Br is
unlikely to be larger than $\approx 10^{-3}$, which is about the branching ratio 
for $B\rightarrow\eta^{\prime}X_{s}$, detection may require $O(10^{10})$ 
B's. Higher luminosity super BELLE/BABAR B-factories as well as efforts at 
hadron B-facilities BTEV and LHCB may well be needed in the hunt for $\chi$.

In the search for $\chi$ the ability to detect small deviations from the SM also 
though requires that we develop tests of the SM that use little or no theory
assumptions and are as free of theory errors as possible. Note in this regard 
that for detection of deviations from the SM at the level of $\approx 10^{-3}$
means that even isospin approximation, widely used in many methods for 
extracting angles of the unitarity triangle can mask $\chi$ and thereby 
defeat the experimental effort for detection of new physics.

Motivated by these considerations we now discuss methods for getting  
unitarity angles with very little theory error, i.e. to $O (< 1\%)$. 

\subsection{Pristine Determination of the Unitarity Triangle via $B \to
K D^0$}. 

Angles of the unitarity triangle can be obtained very
``cleanly" i.e. without any theoretical assumptions from analysis of final 
states
containing $D^0, \bar D^0$ in charged or neutral B-decays\cite{pristine}.

$\gamma$ can be extracted from a study of direct CP violation in
charged B-decays, $B^{\pm} \to K^{\pm} D^0, \bar D^0$ \cite{gronone,atwoodtwo,ygross}. 
$\delta \equiv (\beta -\alpha + 
\pi)=2\beta + \gamma$ \cite{sand,bran,kays,atwoodthree} as  
well as $\beta$ can be obtained from time dependent CP-asymmetry measurements in 
$B^0, \bar B^0 \to K^0 D^0, \bar D^0$.
In both cases common final states of $D^0, \bar D^0$ have to be used,
as flavor tagging of $D^0, \bar D^0$ is very difficult\cite{atwoodtwo}. 
There are 3 types of such common final states:

\begin{enumerate}

\item $D^0, \bar D^0$ decays to CP-non-eigenstates that are doubly Cabibbo 
suppressed \cite{atwoodtwo}, for example,
$K^+ \pi^-$, $K^+ \rho^-$, $K^{*+} \pi^-$ etc.

\item $D^0, \bar D^0$ decays to CP-eigenstates \cite{gronone}, for example, $K^+ 
K^-$,
$\pi^+ \pi^-$, $K^0_S \pi^0$

\item $D^0, \bar D^0$ decays to CP-non-eigenstates that are singly Cabibbo 
suppressed \cite{ygross} for example, $K^{*+} K^{-}$, $K^{-
}K^{*}$, $\rho^{+}\pi^{-}$, $\rho^-\pi^{+}$ etc. 

\end{enumerate}

It turns out that CP asymmetry are expected to be small ($\le 10\%$) for
CPES and for singly Cabibbo suppressed modes (i.e. 2nd and 3rd type) whereas the
interference and CP asymmetry is maximal for CPNES (1st type). On the other 
hand, the branching ratios are expected to be largest for CPES and smallest 
for doubly Cabibbo suppressed CPNES modes.  The general expectations are that 
for 
extraction of $\gamma$, doubly Cabibbo suppressed modes 
should be most efficient among the three types.
However, this is not guaranteed and all three methods should, for sure, be used,  
What is important, for the long run is to note that only two common decay modes 
of $D^0 \bar D^0$ are needed to give enough observables to algebraically solve 
for the CP-odd weak phase $\gamma$, the strong final states phase(s) as well as 
the  suppressed Br($B^- \to K^- \bar{D}^0$) that is very difficult to measure 
experimentally; indeed perhaps a dozen or so such modes are available. This 
should greatly help the analysis in extracting a precise value for $\gamma$ 
without discrete ambiguities.

For time dendent CP asymmetry \cite{atwoodthree} in $B^0, \bar B^0 \to K^0 D^0, 
\bar D^0$,
the discussion is analogous to the above. Again for extraction of
$\delta$ (as well as $\beta$) one needs only two common final states of
$D^0$, $\bar D^0$ from the many; whether they be CPNES, CPES doubly or singly 
Cabibbo 
suppressed
modes, that are available; however, both modes cannot be CPES.

Especially noteworthy is the fact that for clean extraction of the
angles, final states containing $D^0, \bar D^0$ in the decays of
$B^{\pm}, B^0, \bar B^0$ are involved and furthermore common final
states of $D^0,\bar D^0$ decay play a critical role and should aid in increasing
the efficiency of the analysis.

In passing we briefly note of the analogous methods involving
$B_s$ decays to $D_s$ $K^{\pm}$\cite{adk}  (or their 
vector counter-parts\cite{lss}) that can also give $\gamma$
very cleanly\cite{fleischer}.

\section{Exact Chiral Symmetry on the Lattice}
\subsection{Introduction}

In the past few years a significant development for lattice gauge computations 
has taken place.  For the first time, we have practically viable discretization 
methods that exhibit exact chiral symmetry on the lattice even at a finite lattice
spacing, i.e. even before the continuum limit is taken.  By now, not 
only the viability of these methods has been convincingly demonstrated, large 
scale simulations, with some success, have already been using them to address 
some outstanding problem in weak interaction phenomenology pertaining to 
$K\rightarrow 2\pi$ that were very difficult to address heretofore, as briefly 
reminded 
below.

\subsection{Difficulties of calculating weak matrix elements on the lattice with 
conventional discretizations}

Recall that conventionally there are two fermion discretizations: Wilson and 
Staggered (or Kogut-Susskind).  Wilson fermion explicitly break chiral 
symmetry whereas conventional staggered fermions while possessing some residual 
chiral symmetry break flavor symmetry.  The difficulties of maintaining chiral 
symmetry on the lattice is enunciated in the form of a no-go 
(Nielsen-Ninomiya\cite{hbniel}) theorem.

Although conventional wisdom says that these symmetries get restored on the 
lattice in the continuum limit, in practice,
in the study of hadronic weak decays, lack of chiral symmetry imposes an 
extremely serious if not an insurmountable limitation.

Lack of chiral symmetry leads to two types of significant 
difficulties:

1)	Precise renormalization of 4-quark operations can become a difficult fine-
tuning problem.  The point is that, in the absence of chiral symmetry 
operators such as $O_{LL}\equiv[\bar{s}\gamma_{\mu}(1-\gamma_{5})d]^{2}$, 
that are relevant to $K-\bar{K}$ oscillations and $B_{K}$ computation, mix 
under renormalization with wrong chirality operators \cite{cbern}, for example, 
with 
$O_{PP}\equiv(\bar{s}\gamma_{5}d)^{2}$.

The problem is that whereas $<K|O_{LL}|\bar{K}>$ is proportional to the quark 
mass and therefore vanishes in the chiral limit, $<K|O_{PP}|\bar{K}>$ goes to a 
constant in the chiral limit.  Thus even if the mixing coefficients of the wrong 
chirally operators are small you need to know them very accurately in order 
to 
precisely extract the matrix element of physical interest.

2) Mixing with lower dimensional operators is even a worse problem.  This 
happens, for instance, when one considers the operators of the $\Delta S=1$ 
Hamiltonian (e.g. $\bar{s}\gamma_{\mu}(1-\gamma_{5})u\bar{u}\gamma_\mu
(1-\gamma_{5})d$) relevant to $K\rightarrow 2\pi$.  Now such a dim-6 operation 
mixes with 
lower dimensional operations, for example, $\bar{s}d$, $\bar{s}\gamma_{5}d$, 
$\bar{s}\sigma_{\mu\nu} d G^{\mu\nu}\cite{npb262331}$.  The mixing coefficients are now 
power 
divergent, 
for 
example $\sim a^{-n}$ (n=3 for $\bar{s} d$ and $\bar{s}\gamma_{5}d$ and n = 1 
for $\bar{s}\sigma_{\mu\nu} d G^{\mu\nu})$.  So they become increasingly important 
in 
the continuum limit.  Non-perturbative methods (that respect chiral symmetry) 
are essential for handling them.  This was the main reason that early efforts
\cite{bernprl} to 
calculate $\Delta I = \frac{1}{2}$, $K\rightarrow 2{\pi}$ amplitudes on the 
lattice did not make much progress.

\subsection{Domain Wall Fermions}
In 1992, Kaplan \cite{dkapl} in a celebrated paper showed a simple method to 
attain exact 
chiral symmetry on the lattice even at finite lattice spacing.  This remarkable 
feat is accomplished by embedding the 4-dim theory on to 5-dim with a fermion 
mass-term that has the shape of a domain wall across the 4-dimensional boundary 
and switches sign.  The low lying mode bound to the walls then possesses exact 
chiral symmetry on the lattice in the limit that the length $(L_{S})$ of the 
5th dimension has an infinite extent\cite{shamir}.  Nielsen-Ninomiya theorem \cite{hbniel} is 
evaded as 
in Kaplan's construction the number of fermionic degrees of freedom per (4-dim) 
site is no longer finite as assumed in the theorem, and in fact becomes infinite 
as $L_{S}\rightarrow \infty$.  Narayanan and Neuberger \cite{nn} gave an 
elegant flavor 
interpretation of this fifth dimension.

In 1997 1st (quenched) lattice QCD simulation done to test the practical 
viability of this approach showed very encouraging results \cite{tblum}.  In 
those numerical 
simulations for QCD actually the domain wall formulation of Shamir \cite{shamir} 
was  
used.  These early results showed that even with a modest extent of 
the 5th dimension, domain wall fermion possesses very good chiral behavior 
setting 
the stage for their use in large scale simulations.

Since DWF are continuum like their renormalization (perturbative and non-
perturbative) properties are fairly simple.  Also discretization errors tend to 
go as $O(a^{2})$ rendering them with very good scaling properties which tends to 
offset the cost of the extra-dimension.

Since in practice the extent of the 5th dimension is finite, the coupling 
between 
the two walls separated by $L_{S}$ causes a coupling between the light modes and 
gives them a residual mass, $m_{res}$.  This mass can be measured quite 
precisely \cite{blumtwo,ali}.  In low energy applications one can systematically 
include the effect 
of $m_{res}$ in the context of an effective chiral lagrangian \cite{blumthree}.

\subsection{Application to $B_{K}$}

CP-PACS \cite{alikh} and RBC\cite{blumone} collaborations have made   
considerable progress towards a precise calculation of $B_{K}$ with DWF.  Both 
results are in the range of 1-2$\sigma$ below the old result from JLQCD \cite{aokione}, 
$B_{K}$[2 GeV]=$0.628\pm0.042$.  CP-PACS and RBC central values for $B_{K}$ differ 
by about 5-10\%; most likely this difference is due to the fact that CP-PACS 
uses 1-loop lattice perturbation \cite{aokitwo,tblumtwo} theory for 
renormalization of the $\Delta S=2$ operators whereas RBC is using 
non-perturbative renormalization \cite{blumthree}.  
Efforts are now underway to 
repeat this calculation at weaker 
(quenched) coupling \cite{noakitwo} as well as 
with dynamical domain wall quarks \cite{izubuchi}.


\section{$\Delta I=1/2$ Rule and $\epe$: Progress and Outlook.}

\subsection{Introduction}

There have been two recent attempts by the CP-PACS \cite{noaki} and the
RBC \cite{blumone} collaborations, at attacking this old problem on the
lattice using the relatively new discretization method of domain wall
fermions (DWF) \cite{dkapl,shamir,nn}.

First lattice studies of $K\to\pi\pi$ by both CP-PACS \cite{noaki} and
RBC \cite{blumone} with DWF used the lowest order chiral perturbation
theory (ChPT) approach suggested by Bernard \etal \cite{prd322343}.  The
method then calls for using the lattice to compute matrix element of
4-quark operators between $K\to\pi$ and $K\to{}$vacuum which are used
to obtain the corresponding desired $K\to 2\pi$ 
matrix elements\cite{stag}. While
this is a simple method which avoids technical (Maiani-Testa Theorem
\cite{plb245585}) and also practical, computational limitations, it is
nevertheless a severe approximation. In particular, at the leading
order in ChPT being used, final state interactions, which in reality are
very likely important \cite{npb592294} are necessarily absent.  Note
also that in this approximation the good chiral behavior of domain wall
fermions becomes crucial. For one thing in the absence of chiral symmetry, the
unphysical (cubically divergent) contribution from mixing with lower
dimensional operators to $\bra \pi |\theta^{8,1}|K\ket$ cannot be
subtracted away in a relatively simple way by using
$\bra0|\theta^{8,1}|K^0\ket$. Furthermore, the renormalization of
4-quark operators also becomes vastly more complicated due to the
mixing of operators with the wrong chirality ones\cite{cbern}. Since
the $K\to\pi$ matrix elements of some of the operators go to a constant
in the chiral limit, whereas those of the right chirality tend to
vanish, subtraction of the unwanted contribution needs to be done at a
very high precision, i.e.\ it becomes a fine tuning problem. For these
reasons, as mentioned above,  earlier efforts \cite{bernprl} for computation of $K\to2\pi$
and $\epe$ to the
LO in ChPT by the use of $K\to\pi$ and $K\to0$ on the lattice with
Wilson fermions, which explicitly break chiral symmetry, were able to make little headway.

Since chiral symmetry is so critical in the calculation of matrix
elements for $K\to2\pi$ and since, due to the finite extent of the 5th
dimension, rigorously speaking, domain wall quarks do not possess exact
chiral symmetry, it is important to be able to take into account
residual symmetry breaking effects in a systematic fashion.  For matrix
element dominated by long-distance physics this can be accomplished by
shifting the bare masses in ChPT by  $m_{res}$, where $m_{res}$ is the
residual quark mass which the massless quarks on the lattice possess
due to the coupling
between the walls of the 5th dimension \cite{blumthree}. On the lattice
we can calculate $m_{res}$ quite precisely \cite{blumtwo,ali} and the
chiral limit is then taken by setting $(m_{quark} + m_{res})\to0$.

For operators such as $Q_6$ which receive power divergent (i.e.\ short
distance) contributions that are not physical and have to be subtracted
away, the symmetry breaking effect cannot be precisely described by
$m_{res}$ \cite{blumone}. Fortunately, the LEC can still be computed
accurately by taking the slope of the matrix elements with the
$m_{quark}$ so long as $m_{res}$ is independent of $m_{quark}$ to a good 
approximation \cite{blumone}; of course this does require that in actual 
simulations
the length of the 5th dimension is sufficiently long that $m_{res}<<
m_{quark}$.

Note also that for power divergent subtractions ChPT is taken into account to
all orders \cite{blumone}.

Table~\ref{tabnine} gives the (subtracted) $K\to\pi$ matrix element of all of the
4-quark operators of interest for the $I=1/2$ and the $I=3/2$ channel.
This can be used to obtain the $K\to2\pi$ amplitudes via eq.~(201) of
Ref.~\cite{blumone}.

\begin{table}[htb]
\caption{The lattice values for the low energy, chiral perturbation
theory constants decomposed by isospin for $Q_1$ to $Q_{10}$.  (Taken from 
\cite{blumone})
\label{tabnine}}
\begin{center}
\begin{tabular}{lcr}
\hline
\hline
$i$  \qquad\qquad\qquad & $a^{(1/2)}_{i,{\rm lat}}$ &
$a^{(3/2)}_{i,{\rm lat}}$ \\
\hline
1 &  \qquad\qquad $-1.19(31)\times10^{-5}$ \qquad\qquad\qquad &
$-1.38(6)\times10^{-6}$ \\
2 & $2.22(16)\times10^{-5}$ & $-1.38(6) \times10^{-6}$ \\
3 & $0.15(113) \times10^{-5}$ & 0.0 \\
4 & $3.55(96) \times10^{-5}$ & 0.0 \\
5 & $-2.97(100) \times10^{-5}$ & 0.0 \\
6 & $-8.12(98) \times10^{-5}$ & 0.0 \\
7 & $-3.22(16) \times10^{-6}$ & $-1.61(8) \times10^{-6}$ \\
8 & $-9.92(54) \times10^{-6}$ & $-4.96(27) \times10^{-6}$  \\
9 & $-1.85(16) \times10^{-5}$ & $-2.07(9) \times10^{-6}$ \\
10 & $1.55(31) \times10^{-5}$ & $-2.07(9) \times10^{-6}$ \\
\hline\hline
\end{tabular}
\end{center}
\end{table}

Table~\ref{tabten} gives the full results for Re$A_0$, Re$A_2$,
$\omega^{-1}\equiv \frac{{\rm Re}A_0}{{\rm Re}A_2}$ 
and $\epe$ of RBC\cite{blumone}.  

\begin{table}[htb]
\caption {Final values for physical quantities using 
1-loop full QCD
extrapolations to the physical kaon mass and a value of
$\mu=2.13$ GeV for the matching between the lattice and continuum; Taken from \cite{blumone}.  
(Statistical errors only \label{tabten})}.  
\begin{center}
\begin{tabular}{l|c|c}
\hline
\hline
& & This calculation \\
Quantity & Experiment & (statistical errors only) \\
\hline
Re$A_0$(GeV) & $3.33\times10^{-7}$ & $(2.96\pm0.17)\times10^{-7}$ \\
Re$A_2$(GeV & $1.50\times 10^{-8}$ & $(1.172\pm0.053)\times 10^{-8}$ \\
$\omega^{-1}$ & 22.2 & $(25.3\pm1.8)$ \\
Re$(\epe)$ & $(15.3\pm2.6)\times 10^{-4}$ (NA48) & $(-4.0\pm2.3) \times
10^{-4}$ \\
& $(20.7\pm2.8)\times 10^{-4}$ (KTEV) & \\
\hline\hline
\end{tabular}
\end{center}
\end{table}

While both the groups \cite{noaki,blumone} use L0ChPT, DWF and the quenched
approximation, there are some important differences in these two
calculations as well. For one thing, RBC \cite{blumone} used the standard Wilson 
gauge
actions whereas CP-PACS \cite{noaki} used renormalization group improved 
(Iwasaki)
gauge action \cite{iwasaki}. Also in their extractions of Re$A_2$ (and
$B_K$), RBC
used the 1-loop quenched chiral perturbation theory \cite{jhep08023} to
fit
$\bra\pi|Q^{3/2}_{1,2}|K\ket$ whereas CP-PACS used a phenomenological
fit.

\subsection{When the Dust Settles}

\begin{description}

\item[I.] Regarding Re $\epe$

\begin{description}

\item[a)] Key Contributions.

Listed below are the key contributions to  $\epe$ from $I=0$ and 2
final states, all given in units of $10^{-4}$ resulting from\cite{blumone}. Recall, experiment finds
(in this unit) \cite{prl8322,epjc22231} Re $\epe=17\pm2$. (Note
contributions not shown are negligible in comparison).

\begin{center}
\begin{tabular}{lll}
Operator \qquad\qquad  & ~~$I=0$ & ~~$I=2$ \\ \\
$Q_4$ &  $-4.8\pm1.1$ \qquad\qquad & \\
$Q_6$ & $~14.2\pm1.9$ & \\
$Q_8$ & $~1.48\pm.12$ & $-16.97\pm.84$ \\
$Q_9$ & & $~1.56\pm.00$
\end{tabular}
\end{center}

\item[b)] $Q_6$ and $Q_8$ are not the only ones that matter.

Although, as widely expected \cite{rmp7265}--\cite{bardeen}, $Q_6$
and $Q_8$
are the dominant players, due to the cancellations between these two
contributions, other operators (e.g.\ $Q_4$) seem to be making an
appreciable difference to the final result.

\item[c)] Buras approximate formula \cite{hep9908395}.

In this context recall Buras' approximate formula

\be
\epe \simeq \epe |_{6+8} \equiv \epe|_6 + \epe|_8 \label{buras}
\ee

\noi From our lattice data one can see that the contribution of
operators other than $Q_6$ and $Q_8$ is about 60\% of $\epe|_{6+8}$.

\item[d)] Cancellation {\it not\/} between large numbers.

While there is a cancellation between $Q_6$ and $Q_8$, in magnitude
each of this contribution is comparable to the experimental number for
$\epe$. Had it been that

\be
\epe|_6, \quad \epe|_8 >> \epe|_{expt}. \label{cancel}
\ee

\noi then the cancellation would have been between ``large numbers''
(compared to the final result that one is seeking)
and the prognosis for future improvements would have been even harder.

\item[e)] Unnatural Cancellations.

The substantial cancellation $(\sim 85\%)$ between contributions of 
$Q_{6}$ and $Q_{8}$ to $\epsilon'$/$\epsilon$, in all 
likelihood, is 
not natural, i.e. not stable to perturbations.  Recall that these numbers emerge 
after using at least 3 key (uncontrolled) approximations: lowest order chiral 
perturbation theory, quench approximation and heavy charm quark.  It is 
virtually impossible that these approximations affect $Q_{6}$ and 
$Q_{8}$ in the same way.  Indeed there are good reasons to think that both 
chiral perturbation theory and quench approximation are having a bigger effect 
on $Q_{6}$ than on $Q_{8}$.  It seems reasonable therefore to expect 
that in improved calculations of $\epsilon'$/$\epsilon$ these cancellations 
between $Q_6$ and $Q_8$ will not remain.

\item[f)] Phenomenological bound.

To the extent that $\epe|_{I=2}<0$, there is a useful pheomenological
bound,

\be
\epe|_{I=0} > \epe|_{expt} \label{bound}
\ee

\noi with which one can test the SM\null. For the purpose of this test
the cancellations between the $I=0$ and $I=2$ contributions are not
quite relevant.  Note that our current data gives left hand side of
eqn.(16) of $\sim11\pm2$ whereas the RHS (from experiment) is
$\sim17\pm2$. Clearly then $\epe|_{I=0}$ must increase appreciably as
improvements in our lattice calculation are made if the SM's
description of CP is to continue to hold.

These considerations suggest that tests of the SM with improvements in
accuracy appear feasible.

\end{description}

\item[II)] Repercussions for the origin of the $\Delta I=1/2$ Rule.

The octet enhancement, i.e.\ $\frac{{\rm Re}A_0}{{\rm Re}A_2} \sim 20
>>1$, has been a long standing puzzle in Particle Physics. The lattice
calculation with domain wall quarks, although not having sufficient
control over all the systematic errors cannot at present give a reliable result for
$\epe$, they do provide with a useful and unambiguous information for the
$\Delta I=1/2$ enhancement.  The lattice result leads to an important
and remarkable conclusion regarding the $\Delta I=1/2$ rule as can be
seen from Table~\ref{tabnine}: The contribution of Re$A_2$ and
especially
of Re$A_0$ originate almost entirely from the aboriginal 4-fermi
operator $Q_2$. Indeed, we find

\begin{center}
\begin{tabular}{lcc}
Operator & \qquad Re$A_0$ (GeV) \qquad  & Re$A_2$ (GeV) \\ \\
$Q_1$ & \qquad $(3.48\pm.77)\times10^{-8}$ \qquad &
$(-.363\pm.016)\times10^{-8}$ \\
$Q_2$ & \qquad $(24.5\pm1.6)\times10^{-8}$ \qquad  &
$(1.520\pm.068)\times10^{-8
}$ \\
$Q_6$ & \qquad $(0.050\pm0.006)\times10^{-8}$ \qquad  & \\
\end{tabular}
\end{center}

\noi These numbers should be compared to the experimental ones:
Re$A_0^{expt}=33.30\times10^{-8}$ GeV, Re$A_2^{expt}=1.50\times10^{-8}$
GeV\null. Clearly Re$A_0$ is completely dominated by $Q_2$, making about
$\sim 80$--85\% contribution and $Q_1$ makes the remaining $\sim 15\%$
contribution to Re$A_0$.  In particular the contribution of $Q_6$ to
Re$A_0$ is completely negligible, being $\sim0.2\%$.  This is in
sharp contrast to some model calculations for $\epe$ in which contribution of
$Q_6$ to Re$A_0$ is typically almost 20--30\%
\cite{rmp7265,hep0001235}. Note that while numbers given here for
contribution of individual operators are based on calculations at
$\mu\sim2$ GeV we have
studied the $\mu$ dependence from $\sim 1.3$--2 GeV and the dependence
is quite mild \cite{blumone}.

It must be emphasized that all these quantitative findings, in
particular $Q_6$ vs $Q_2$ contributions to $K\to 2\pi(I=0)$, are based
on LOChPT calculations and this could change as higher order
corrections are included.

It is indeed interesting and ironical that the penguin operators
\cite{npb120316} originally invoked to explain the $\Delta I=1/2$ rule
seem to play little role therein, at least in the context 
of our lattice calculation\cite{blumone}. However, the subsequent conjecture of
their importance to rendering a largish $\epe$ \cite{prd202392} seems
to be substantiated although significant theoretical progress still
needs to be made before the repercussions of the precise experimental
measurement can be fully assessed.

\end{description}

\subsection{Approximations and Concerns}

There were several approximations made in the lattice of
calculations \cite{noaki,blumone} using DWF that were recently completed.

\begin{enumerate}

\item The quenched approximation so that quark antiquark loops are
ignored in the propagation of the gauge field (gluon).

\item Lowest order chiral perturbation theory (LOChPT), so that the
matrix elements of the 4-quark operators are calculated in the leading
order (LO) in this approximation. This means that operators
[$Q_1$--$Q_6$, $Q_9$, $Q_{10}$] that transforms as (8,1) and/or (27,1)
under SU(3)$_L\times$SU(3)$_R$ are calculated to $O(p^2)$ whereas those
of $Q_7$, $Q_8$ which transform as (8,8) are to $O(p^0\to1)$ in the
chiral expansion.

\item The charm quark is assumed to be very heavy and  
integrated out.  The effective Hamiltonian \cite{npb408209,zpc68239},
consequently consists of only 3 active flavors: $u$, $d$, $s$.

\end{enumerate}

These approximations are uncontrolled, i.e.\ we do not have a reliable
estimate of  how inaccurate they are. While the quenched approximation
seems to be accurate to 10--15\% in many spectrums and decay constant
calculations, it may well be a lot worse for some hadronic matrix
elements. In particular, comparison of the analytical formulas for $\bra
\pi|Q^{3/2}_{7,8}|K\ket$ to NLO \cite{hep9912513} in full ChPT with the
lattice data \cite{blumone} obtained using the quenched approximation
shows that the logs of the full ChPT seem to be absent \cite{prd}. Also
the matrix element of $Q_6$, which is of crucial importance to $\epe$
is claimed to be very susceptible to quenching effects
\cite{hep0108029}.

Although, in many low energy applications, LOChPT works fairly well,
in $K\to2\pi$ there are reasons to be suspicious. First of all an
important mass scale here is $m_K$ and not just $m_\pi$. Furthermore,
in the $I=0$ channel the $\pi$-$\pi$ rescattering effects (FSI), which
cannot occur in the LOChPT, are likely to be quite important
\cite{npb592294}. Indeed for $\bra\pi\pi|Q_6|K\ket$ higher order chiral
corrections may well be intertwined with a $O^{++}(\sigma)$ resonance
in the $\pi$-$\pi$ channel \cite{isgur,cnpp20119}.

Since the mass of the charm quark is only $\sim 1.3$ GeV, integrating it
out assuming it is very heavy (i.e.\ $>>\Lambda_{QCD}$) is very likely
not a
good approximation. Corrections from higher dimensional operators are
likely to
be sizeable \cite{jhep0010048}. Also in the 4-flavor theory, GIM
cancellation forbids power-divergent mixing of dim-6 operators of the
$I=0$, $H_{eff}$ with lower dim operators, so the 4 flavor theory is
preferable over the 3 flavor theory for that reason too, although this
advantage is only relevant in computation of Re$A_0$
\cite{npb289505}--\cite{hep0011070}. In the calculation of Im$A_0$
where (8,1) penguin operators, such as $Q_6$, become important which
originate from integrating out the top quark, the top quark is so heavy
compared to the lattice cut-off that integrating it out is
unavoidable.

\subsection{Future Outlook}

While the above approximations used in the current lattice calculations
are not controllable, systematic improvements are feasible and
efforts at these are well underway.

First, recent studies of renormalization group-improved gauge actions in
the context of domain wall fermions appear very promising in significant
improvements in chiral symmetry (which was already remarkably good)
\cite{hep0110074}. Efforts are also underway towards creating a large
ensemble of gauge configurations with dynamical (2 flavor) domain wall
quarks \cite{izubuchi} with lattice spacing $\sim(2$ GeV$^{-1})$. This
should allow a first study of the quenching effects in $K\to2\pi$ matrix
elements in another year or so. Also work is being done at finer
lattice spacing $\sim (3$ GeV$^{-1})$ with the hope that this will allow
a
better treatment of the charm-quark and a calculation of the matrix
elements in the effective theory with 4 active flavors ($u$, $d$, $s$,
$c$) \cite{noakitwo}.

Note also that new calculations of the $K\to2\pi$ matrix elements have
begun \cite{giusti} using another discretization (overlap fermions
\cite{neuberger}) possessing excellent chiral symmetry.

Recent works also show how lattice computations of all the matrix
elements relevant to $K\rightarrow 2{\pi}$ and $\epsilon'/\epsilon$ can be 
obtained beyond the leading order in ChPT. In one method
\cite{prd} matrix elements of all the relevant operators ($\Delta
I=1/2$ or 3/2) can be obtained to NLO by using lattice computations of
$K$-$\bar K$, $K\to\pi$, $K\to0$ and $K\to2\pi$ at the two unphysical
kinematics ($m_K=m_\pi$ and $m_K=2m_\pi$) wherein Maiani-Testa Theorem
\cite{plb245585}
can be evaded. In another construction \cite{hep0110206} the $K\to2\pi$
matrix elements for $\Delta I=\frac{3}{2}$ transitions can be obtained to NLO by 
using lattice computations of
$K\to2\pi$ with momentum insertion on one of the final state pions.

Indeed in a very interesting paper Lellouch and Luscher have also
proposed a method wherein $K\to2\pi$ matrix elements may be directly
calculated without using ChPT by relating them to finite volume
correlation functions \cite{hep0003023}. There is also a proposed 
method which makes use of dispersion relations to calculate
physical $K \to \pi \pi$ amplitudes to all orders
in ChPT\cite{disp}. We note that both of these methods make rather stringent demands
on unitarity therefore their implementation, especially for the
$\Delta I=1/2$ case, may need full QCD simulations. 

\section{Unitarity Triangle from $K$-Decays}

While determination of the unitarity triangle (UT)
from $B$ decays has been receiving considerable attention and is much
in the news, it is useful to note that not only an independent
determination of the UT can also be made purely from $K$-decays but it
is important to do so. In principle, 
there are four physical processes that can 
be
useful here, whereas any three of them would be sufficient.

\begin{enumerate}

\item $\epsilon$, the indirect CP-violation parameter characterizing the
CP-violation  in $K_L\to2\pi$.

\item The $Br(K^+\to\pi^+\nu\bar\nu)$, a determination of which has been
underway for a long time and a crude measurement now exists thanks to
the two candidate events that have been seen, $Br(K^+\to\pi^+\nu\bar
\nu) = (1.57^{+1.75}_{-.82})\times10^{-10}$ \cite{e787}.

\item There is considerable experimental interest in measuring the
$Br(K_L\to\pi^0\nu\bar\nu)$. This is a very interesting mode which is
CP violating \cite{llitt} and is theoretically extremely clean. 
but
clearly an
extremely difficult experimental challenge.

\item The direct CP-violation parameter $\epe$. Although the
experimental number is now quite precisely known
\cite{prl8322,epjc22231}, it can only be useful in the context of the
UT, if the theory can be brought under-control.  Renewed interest on
the lattice, in light of recent progress in maintenance of chiral
symmetry on the lattice (described briefly in preceding pages) gives one some 
encouragement that
perhaps
a few
years down the road we would be able to make use of the experimental
result and translate it into the CP violation parameter $\eta$ of the CKM 
paradigm.
In the $\rho$-$\eta$ plane, $\epe$, when the numerical value of $\ep$
is taken from experiment, would provide a horizontal line (actually a
band due to the error in theory and in experiment).

As is well known transplanting $\epsilon$ to $\rho$, $\eta$ plane does
require knowledge of the non-perturbative hadronic parameter, $B_K$.
Fortunately as already mentioned, lattice calculations of $B_K$ are now
quite mature. In fact several different discretization methods have
been used to determine this important quantity. While the current
accuracy is around 15\%, efforts with dynamical quarks are underway and
in 3--5  years we should expect the accuracy to improve appreciably.

\end{enumerate}

The theory for $K^+\to\pi^+\nu\bar\nu$ is also rather clean \cite{hep0101336}.
The basic process $s\to d\nu\bar\nu$ is dominated by the top quark.
Conversion of that to $K^+\to\pi^+\nu\bar\nu$ can be done using isospin
by relating it to the thoroughly studied charge current process $K\to
\pi e\nu$.

The observed $Br$, deduced on the basis of the 2 events seen so far, is
consistent with the expectation from the SM; our
\cite{atwoodone} CKM fits give,
$BR(K^+\to\pi^+\nu\bar\nu)=(0.67\pm0.10)\times 10^{-10}$. Experimental
efforts are underway to improve this measurement in the near future at
BNL and further down the road at FNAL \cite{skett}\null.

The decay $K^0_L\to\pi^0\nu\bar\nu$ is fascinating as it is CP
violating. The theory \cite{hep0101336} in this case is even clearer
then for the charged counter part and  $BR(K^0_L\to K^0\nu\bar\nu) =
1.5\times 10^{-3} A^4 \lambda^{10}\eta^2$. Our \cite{atwoodone} fit
value
is $BR(K_L\to\pi^0 \nu\bar\nu) = (.23\pm.07)\times 10^{-10}$. So an
experimental measurement would give a clean determination of the CKM
phase $\eta$. Note though that $(A\approx V_{cb})$ the current
accuracy  in $A(\sim7\%)$ should be improved otherwise it introduces
significant error on the $\eta$ determination. The
$K_L\to\pi^0\nu\bar\nu$ experiment is clearly very challenging and it
is receiving attention at KEK (E391), at BNL (KOPIO) and at FNAL
(CKM)\cite{skett}.

Given the intrinsic difficulties of this experiment and those of an
accurate theoretical calculation of $\epe$ it would be interesting to
see
which of these is brought under control first.
\bigskip

{\bf Acknowledgements} \\
I thank the organizers for inviting me and for the useful
and enjoyable meeting. This research was supported in part by
USDOE grant \# DE-AC02-98CH10886.


\begin{thebibliography}{99}


\bibitem{ckm}

N. Cabibbo, Phys. Rev. Lett {\bf10},
531 (1963); M. Kobayashi and T. Maskawa, Prog. Theo. Phys. {\bf 49}, 652
(1973).  

\bibitem{pristine} D. Atwood and A. Soni, hep-ph/021207.

\bibitem{dwein} D. Weingarten, Phys. Lett.{\bf B99},333 (1982); H. Hamber 
and G. Parisi, Phys. Rev. Lett. {\bf 47}, 1792 (1981).

\bibitem{dtous} D. Toussaint, Nucl. Phys. {\bf B106} (Proc. Suppl.), 111 
(2002); T. Kaneko, {\it ibid} 133.

\bibitem{heplat} C. Bernard {\it \etal}(MILC), hep-lat/0209163.

\bibitem{aokione} S. Aoki {\it \etal} (JLQCD Collab.), Phys. Rev. Lett.
{\bf 80}, 5271 (1998).   

\bibitem{kilcup} G. Kilcup, R. Gupta and S.R. Sharpe, Phys. Rev. {\bf
D57}, 1654 (1998).  

\bibitem{cbas88} C. Bernard, A. Soni,
Nucl.Phys.Proc.Suppl.9:155-174,1989. 

\bibitem{ssharp} S. Sharpe, Nucl. Phys. {\bf B34} (Proc. Suppl.), 403 
(1994).

\bibitem{becir} D. Becirevic {\it \etal} (SPQCDR Collab.),
hep-lat/0209136. 

\bibitem{tblum} T. Blum and A. Soni, Phys. Rev. {\bf D56}, 174
(1997); Phys. Rev. Lett. {\bf 79}, 3595 (1997).  

\bibitem{alikh} A. Ali Khan {\it \etal} (CP-PACS Collab.), Phys. Rev. {\bf
D64}, 114506 (2001).   

\bibitem{blumone} T. Blum {\it \etal} (RBC Collab.), hep-lat/0110075.   

\bibitem{milc} T. DeGrand (MILC), hep-lat/0208054.

\bibitem{ngarr} N. Garron {\it \etal}, hep-lat/0212015.

\bibitem{izubuchi} T. Izubuchi {\it \etal} (RBC Collab.), hep-lat/0210011.  

\bibitem{bls} C. Bernard, J. Labrenz and A. Soni, Phys. Rev. {bf D49}, 
2536(1994).

\bibitem{nyam} N. Yamada, Nucl. Phys. {\bf B119} (Proc. Suppl.),93 (2003).

\bibitem{nucl} A. Soni, Nucl. Phys. {\bf B47} (Proc. Suppl.), 43(1996).

\bibitem{hepph} A. Soni, hep-ph/0003092.

\bibitem{co94} C. Bernard, {\it \etal} (MILC), {\it Phys. Rev.} {\bf 
D66},094501(2002).

\bibitem{error} In quoting this error for Refs.\cite{heplat}
and \cite{co94}, I have 
taken the liberty to add linearly the error due to chiral extzapolertion to the 
other errors which are (as usual) added in quadrature.

\bibitem{noaki} J.I. Noaki \etal [CP-PACS Collab.], hep-lat/0108013.

\bibitem{hep0101336} See, e.g., A. J. Buras, hep-ph/0101336.   

\bibitem{kpart} K. Hagiwara {\it \etal} (Particle Data Group Collaboration),  
Phys. Rev. {\bf D66},010001(2002).

\bibitem{lepb} For LEP see, 
http://lepborc.wel.cern.
ch/LEPBOSC/LEPBOSC98/3; for SLD see K.Abe {\it \etal},  hep-ex/0012043;
hep-ex/0011041.

\bibitem{bern} C. Bernard, T. Blum and A. Soni, Phys.\ Rev.\ {\bf D58},
01401 (1998).   

\bibitem{buch} G. Buchalla {\it \etal}  Rev. Mod. Phys. {\bf 68}, 1125
(1996).  

\bibitem{tinam} T. Inami and C.S. Lim, Prog. Theor. Phys. {\bf 
65},1772,1981.

\bibitem{aelk} See e.g. A. El-Khadra {\it \etal}, Phys. Rev., {\bf D64}, 
014502 (2001).

\bibitem{atwoodone} D. Atwood and A. Soni,  hep-ph/0103197.   

\bibitem{ciuch} M. Ciuchini {\it \etal} hep-ph/0012308.    

\bibitem{hocker} A. Hocker {\it \etal} hep-ph/0104062.   

\bibitem{intmor} See, e.g. talks given by A. Soni at the Hadron Phenomenology 
workshop at INT Seattle, 11/8/01 (http://mocha.phys.washington.edu/~ 
int\_talk/Worskshops/int\_01\_3/People/Soni\_A)
and at the Electroweak Session of 
XXXVII Rencontres de Moriond, Les Arcs, March 9-16,2002 
(http://moriond.in2p3.fr/EW/2002/transparencies/index.html)

\bibitem{aknon} A. Kronfeld and S. Ryan, hep-ph/0206058.


\bibitem{llellh} L. Lellouch, hep-ph/0211359.

\bibitem{belle} B. Aubert {\it \etal} (BABAR Collab), 
hep-ex/0207042; K. Abe {\it 
\etal} (BELLE Collab) hep-ex/0208025.

\bibitem{ynirh} See, e.g., Y. Nir, hep-ph/0208080.

\bibitem{ashlat} A. Soni, hep-lat/9510036.

\bibitem{dga} D. Atwood, G. Eilam and A. Soni, hep-ph/9409229.

\bibitem{ksw} A. Khodjamirian \etal, hep-ph/9506242; 
Korchemsky \etal, Phys. Rev. {\bf D61},114510,2000;   
Colangelo, F. De Fazio and G. Nardulli, hep-ph/9606219; C.Q. Geng, G. C. Lih and 
W. M. Zhang, hep-ph/9710323; G. Eilam, I. Halperin and R. Mendel, hep-
ph/9506264; S. Descotes-Genon and C. Sachrajda, hep-ph/0209216;
E. Lunghi, D. Pirjol and D. Wyler, hep-ph/0210091; S. Bosch \etal,
hep-ph/0301123.  

\bibitem{hsieh} 
C. Bernard, P. Hsieh and A. Soni, hep-lat/9311010.

\bibitem{bufor} D. Buford {\it \etal} (UKQCD Collab) hep-lat/9503002; A. Abada 
{\it \etal} (APE Collab) hep-lat/9503020; D. Becirevic, hep-ph/0211340.

\bibitem{ahmed} See also, A. Ali and A. Y. Parkhomenko, Eur. Phys. J.
{\bf C23}, 89(2002). 

\bibitem{blak} D. Atwood, B. Blok and A. Soni, hep-ph/9408373.

\bibitem{scidac} http://www.osti.gov/scidac/henp/index.html.

\bibitem{atwhep} D. Atwood {\it \etal}, hep-ph/0006032.

\bibitem{gronone}  M. Gronau and D. Wyler, Phys. Lett. {\bf B265},
172 (1991).  

\bibitem{atwoodtwo} D. Atwood, I. Dunietz and A. Soni, Phys. Rev.
Lett. {\bf78}, 3257 (1997); Phys. Rev. {\bf D63}, 036005 (2001).

\bibitem{ygross} Y. Grossman, G. Ligeti and A. Soffer, hep-ph/0210433.

\bibitem{sand} A.I. Sanda, hep-ph/0108031.   

\bibitem{bran} G. Branco, L. Lavora and J. Silva, in``CP Violation'',
Oxford University Press (1999).   

\bibitem{kays} B. Kayser and D. London, hep-ph/9905561.   

\bibitem{atwoodthree} D. Atwood and A. Soni, hep-ph/0206045.  

\bibitem{adk}
R. Aleksan, I. Dunietz and B. Kayser, Z. Phys. {\bf C54},
653(1992).

\bibitem{lss}
D. London, N. Sinha and R. Sinha, Phys. Rev. Lett.
{\bf 85}, 1807(2000).

\bibitem{fleischer} See also R. Fleischer hep-ph/0301255.

\bibitem{hbniel} H.B. Nielsen and M. Ninomiya, Phys. Lett. {\bf 
B105},219('81).

\bibitem{cbern}  G. Martinelli, Phys. Lett.{\bf B41} 395(94);
C. Bernard, T. Draper and A. Soni, Phys. Rev.{\bf D36}, 
3224(1987). 

\bibitem{npb262331} M. Bochicchio {\it \etal} Nucl. Phys. {\bf B262},
331 (1985).   

\bibitem{bernprl} C. Bernard {\it \etal} Phys. Rev. Lett. {\bf 55},
2770 (1985)   

\bibitem{dkapl} D. Kaplan, hep-lat/9206013.

\bibitem{shamir} Y. Shamir, Nucl.\ Phys.\ {\bf B406}, 90 (1993); V.
Furman and Y. Shamir, Nucl.\ Phys.\ {\bf B439}, 54 (1995).

\bibitem{nn}
R. Narayanan and 
H. Neuberger. Phys. Lett. {\bf B302},62(1993); Nucl. Phys. {\bf 
B443}, 305(1995).

\bibitem{blumtwo} T. Blum {\it \etal} (RBC Collab.), hep-lat/0007038.   

\bibitem{ali} A. Ali Khan {\it \etal} (CP-PACS Collab.), Phys. Rev. {\bf
D63}, 114504 (2001).   

\bibitem{blumthree} T. Blum {\it \etal} (RBC Collab.) Phys. Rev. {\bf D66},
014504 (2002).  

\bibitem{aokitwo} S. Aoki, T. Izubuchi, Y. Kuramashi and Y. Taniguchi,
Phys. Rev. {\bf D59}, 094505 (1999); S. Aoki and Y. Kuramashi, Phys.
Rev. {\bf D63}, 054504 (2001).   

\bibitem{tblumtwo} T. Blum, A. Soni and M. Wingate, Phys. Rev. {\bf
D60}, 114507 (1999).   

\bibitem{noakitwo} J.-I. Noaki {\it \etal} (RBC Collab.)
hep-lat/0211013.  

\bibitem{prd322343} C. Bernard {\it \etal} Phys. Rev. {\bf D32} 2343
(1985).  

\bibitem{stag} For efforts at computing $K \to \pi \pi$
on the lattice using staggered fermions see: D. Pekurovsky and 
G. Kilcup, Phys.\  Rev.\ D {\bf 64},
074502 (2001); T. Bhattacharya, G.T. Fleming, G. Kilcup, R.
Gupta, W. Lee, S. Sharpe, Nucl.Phys. Proc. Suppl. 106 (2002) 311-313.

\bibitem{plb245585} L. Maiani and M. Testa, Phys. Lett. {\bf B245},
585 (1990).  

\bibitem{npb592294} See, e.g. E. Pallante and A. Pich, Nucl. Phys.
{\bf B592}, 294 (2001).  

\bibitem{iwasaki} Y. Iwasaki, unpublished, UTHEP-118(1983).  

\bibitem{jhep08023} M. Golterman and E. Pallante, (JHEP) {\bf08}, 023
(2000).  

\bibitem{prl8322} A. Alavi-Harari {\it \etal} (KTeV Collab.), Phys. Rev.
Lett.{\bf 83}, 22 (1999).  

\bibitem{epjc22231} A. Lai {\it \etal} (NA48 Collab.), Eur. Phys. J. {\bf
C22}, 231 (2001).   

\bibitem{rmp7265} S. Bertolini, M. Fabbrichesi and J.O. Eeg, Rev.
Mod. Phys. {\bf72}, 65 (2000).  

\bibitem{hep0001235} S. Bertolini, hep-ph/0001235.   

\bibitem{hep9908395} A.J. Buras, hep-ph/9908395.   

\bibitem{bardeen} See also, T. Hambye et al, Phys. Rev. {\bf D58},
14017, 1998; J. Bijnens and J. Prades, JHEP 0001, 002, 2000.



\bibitem{npb120316} M.A. Shifman, A.I. Vainshtein and V.I. Zakharov, Nucl. 
Phys. {\bf B120}, 316 (1977).

\bibitem{prd202392} F.J. Gilman and M.B. Wise, Phys. Lett. {\bf B83},
83 (1979); Phys. Rev. {\bf D20}, 392 (1979).  

\bibitem{npb408209} A. Buras, M. Jamin and M.E. Lautenbacher, Nucl.
Phys. {\bf B408}, 209 (1993).   

\bibitem{zpc68239} M. Ciuchini, E. Franco, G. Martinelli, L. Reina and
L. Silvestrini, Z. Phys.\ {\bf C68}, 239 (1995).  

\bibitem{hep9912513} V. Cirigliano and E. Golowich, hep-ph/9912513;
hep-ph/0109265.  

\bibitem{prd} J. Laiho and A. Soni, {\it Phys. Rev.} {\bf D65}, 114020, 2002;
see also hep-lat/0306035.    

\bibitem{hep0108029} M. Golterman and E. Pallante, hep-let/0108029.

\bibitem{isgur} N. Isgur, K. Maltman, J. Weinstein and T. Barnes,
Phys. Rev. Lett. {\bf64}, 161 (1990).   

\bibitem{cnpp20119} U.-G. Meissner, Comments on Nucl. \& Particle
Physics, {\bf 20}, 119 (1991).  

\bibitem{jhep0010048} V. Cirigliano, J. Donoghue and E. Golowich, (JHEP)
{\bf0010}, 048 (2000).  

\bibitem{npb289505} L. Maiani, G. Martinelli, G.C. Rosti and M. Testa,
Nucl. Phys. {\bf B289}, 505 (1987).   

\bibitem{npb4883} C. Bernard, T. Draper, G. Hockney and A. Soni, Nucl.
Phys. B (Proc. Suppl.) {\bf4}, 883 (1988).  

\bibitem{hep0011070}  S. Capitani and L. Giusti, hep-lat/0011070.  

\bibitem{hep0110074} K. Orginos \etal [RBC Collab.], hep-lat/0110074.

\bibitem{giusti}  L. Giusti,hep-lat/0212012.   

\bibitem{neuberger} H. Neuberger, Phys. Lett. {\bf B417}, 141 (1998); 
Phys. Rev. {\bf D57},5417(1998). 

\bibitem{hep0110206} P. Boucaud {\it \etal} (SPQCDR Collab.),hep-lat/0110206; 
C.J.D. Lin, G. Martinelli, E. Pallante, C. Sachrajada and G. Villadoro, hep-
lat/0208007.   

\bibitem{hep0003023} L. Lellouch and M. Luscher, hep-lat/0003023;
C. J. D. Lin \etal, Nucl. Phys. {\bf B619}, 467 (2001).    

\bibitem{disp}
M. Buechler \etal, Phys. Lett. {\bf B521}, 22 (2001).

\bibitem{e787} S. Adler {\it \etal} (E787 Collab.),Phys. Rev.
Lett. {\bf 88}, 041803 (2002).  

\bibitem{llitt} L. Littenberg, Phys. Rev.{\bf D39},3322(1989).

\bibitem{skett} S. Kettell, hep-ex/0205029.
















































\end{thebibliography}
\end{document}